\documentstyle[11pt,aaspp4]{article}

\slugcomment{To appear in the Astronomical Journal (accepted 1999 March 1)}

\lefthead{MIGHELL \& BURKE}
\righthead{URSA MINOR DWARF SPHEROIDAL GALAXY}

\def\noteforeditor#1{}
\def\firstuse#1{}

\lefthead{MIGHELL \& BURKE}
\righthead{URSA MINOR DWARF SPHEROIDAL GALAXY}

\begin{document}

\def\et{\hbox{et~al.\ }}
\def\vs{\hbox{vs }}
\def\feh{\hbox{[Fe/H]}}
\def\bmv{\hbox{$B\!-\!V$}}
\def\vmi{\hbox{$V\!-\!I$}}
\def\ebmv{\hbox{$E(\bmv)$}}
\def\dbmv{\hbox{$d_{(B\!-\!V)}$}}
\def\dbmr{\hbox{$d_{(B\!-\!R)}$}}
\def\mvto{\hbox{$M_V^{\rm{TO}}$}}
\def\mvrhb{\hbox{$M_V^{\rm{RHB}}$}}
\def\vhb{\hbox{$V_{\rm{HB}}$}}
\def\lea{\mathrel{<\kern-1.0em\lower0.9ex\hbox{$\sim$}}}
\def\gea{\mathrel{>\kern-1.0em\lower0.9ex\hbox{$\sim$}}}
\def\ifundefined#1{\expandafter\ifx\csname#1\endcsname\relax}

\title{WFPC2 OBSERVATIONS OF THE URSA MINOR\\
DWARF SPHEROIDAL GALAXY\altaffilmark{1}
}

\author{
{\sc
Kenneth J. Mighell\altaffilmark{2}
and
Christopher J. Burke\altaffilmark{3}\altaffilmark{,4}
}
}
\affil{Kitt Peak National Observatory,
National Optical Astronomy Observatories\altaffilmark{5},\\
P.~O.~Box 26732, Tucson, AZ~~85726 \\ {\it mighell@noao.edu,
christopher.burke@yale.edu}}

\altaffiltext{1}{
Based on observations made with the NASA/ESA
{\sl{Hubble Space Telescope}},
obtained from the data archive at the Space Telescope
Science Institute,
which is operated by the Association of
Universities for Research in Astronomy, Inc.\ under NASA
contract NAS5-26555.
}

\altaffiltext{2}{
Guest User, Canadian Astronomy Data Centre, which is operated by the
Dominion Astrophysical Observatory for the National Research Council of
Canada's Herzberg Institute of Astrophysics.
}

\altaffiltext{3}{Based on research conducted at NOAO as part of the
Research Experiences for Undergraduates program.}

\altaffiltext{4}{{\em{Current address:~}}
Department of Astronomy,
Yale University,
P.~O.~Box 208101,
New Haven, CT~~06520-8101
}

\altaffiltext{5}{NOAO is operated
by the Association of Universities for Research in Astronomy, Inc., under
cooperative agreement with the National Science Foundation.}

\newpage
\begin{abstract}
We present our analysis of archival
{\sl{Hubble Space Telescope}}
Wide Field Planetary Camera 2  (WFPC2) observations in
F555W ($\sim$V) and F814W ($\sim$I) of the central region of the
Ursa Minor dwarf spheroidal galaxy.
The $V$ versus $V-\!I$ color-magnitude diagram
features a sparsely populated blue horizontal branch,
a steep thin red giant branch,
and a narrow subgiant branch.
The main sequence reaches $\sim$2 magnitudes below the
main-sequence turnoff
($V_{\rm{TO}}^{\rm UMi}\approx23.27\pm0.11$ mag)
of the median stellar population.
We compare the fiducial sequence of Ursa Minor
with the fiducial sequence of the
Galactic globular cluster M92 (NGC 6341).
The excellent match between Ursa Minor and M92
confirms that the median stellar population of the UMi dSph galaxy
is metal poor
(${\rm[Fe/H]}_{\rm UMi}
\approx
{\rm[Fe/H]}_{\rm M92}
\approx{\rm-2.2}$ dex)
and ancient
($age_{\rm UMi}
\approx
age_{\rm M92}
\approx14$ Gyr).
The $\bmv$ reddening and the absorption in $V$ are estimated to be
$E(\bmv)=0.03\pm0.01$ mag
and
$A_V^{\rm UMi} = 0.09\pm0.03$ mag.
A new estimate of the distance modulus of Ursa Minor,
$(m-M)_0^{\rm UMi} = 19.18 \pm 0.12$ mag,
has been derived based on fiducial-sequence fitting with M92
[$\Delta V_{\rm{UMi-M92}} = 4.60\pm0.03$ mag
and $\Delta (\vmi)_{\rm{UMi-M92}} = 0.010\pm0.005$ mag]
and the adoption of the apparent $V$ distance modulus for M92
of $(m-M)_V^{\rm M92} = 14.67\pm0.08$ mag
(Pont \et \cite{poet1998}, \aap, 329, 87).
The Ursa Minor dwarf spheroidal galaxy is then at a distance of
$69\pm4$ kpc from the Sun.
These {\sl{HST}} observations indicate that Ursa Minor has had a
very simple star formation
history consisting mainly of a single major burst
of star formation about 14 Gyr ago
which lasted $\lea$2 Gyr.
While we may have missed minor younger stellar populations
due to the small field-of-view of the WFPC2 instrument,
these observations clearly show that most of the
stars in the central region
Ursa Minor dwarf spheroidal galaxy are ancient.
If the ancient Galactic globular clusters, like M92, formed
concurrently with the early formation of 
the
Milky Way galaxy itself,
then the Ursa Minor dwarf spheroidal is probably as old as the Milky Way.
\end{abstract}

\keywords{
galaxies: abundances
---
galaxies: evolution
--
galaxies: individual (Ursa Minor)
--
Local Group
}

\clearpage
\section{INTRODUCTION}

The Ursa Minor (UMi) dwarf spheroidal (dSph) galaxy
was independently discovered by
Wilson (\cite{wilson1955})
and
Hubble.
Ursa Minor is the second closest satellite galaxy of the Milky Way
at a distance of $69\!\pm\!4$ kpc ($\sim$220,000 light years) from the Sun.
Color-magnitude diagrams of the brightest stars of this
faint
($M_V\!\approx\!-8.9$ mag:
Kleyna \et \cite{kleyna_etal_1998})
small
(
$r_{\rm{tidal}}=628\!\pm\!74$ pc:
Irwin \& Hatzidimitriou \cite{irha1995})
galaxy feature a strong blue horizontal branch
(e.g.,
van Agt \cite{vanagt1967};
Cudworth, Olszewski, \& Schommer \cite{cuolsc1986};
Kleyna, \et \cite{kleyna_etal_1998}) --- a unique horizontal branch morphology
amongst the nine Galactic dSph satellite galaxies.
The deep $BV$ CCD observations of
Olszewski \& Aaronson (\cite{olaa1985})
indicate that Ursa Minor has an age and abundance very similar to that of
the ancient metal-poor Galactic globular cluster M92 (NGC 6341).
Ursa Minor may be the only dwarf galaxy in the Local Group which is composed
exclusively of stars older than 10 Gyr (Mateo \cite{mateo1998}).

In this work we investigate the star formation history of the
Ursa Minor spheroidal galaxy using archival
{\sl Hubble Space Telescope} WFPC2 data.
Section 2 is a discussion of the observations and photometric reductions.
We present and compare our results with previous work
in Sec.\ 3. The paper is summarized in Sec.\ 4.
Appendix A describes a new robust algorithm for the computation of
of fiducial sequences from high-quality stellar photometry.

\section{OBSERVATIONS AND PHOTOMETRY}

The Ursa Minor dwarf spheroidal galaxy was observed with the
{\sl{Hubble Space Telescope}} ({\sl{HST}})
Wide Field Planetary Camera 2 (WFPC2) on
1995 July 4 through the F555W ($\sim$$V$) and F814W ($\sim$$I$)
filters.
The WFPC2 WFALL aperture
(Biretta \et \cite{biet1996})
was centered on the target position
given in
Table\ \ref{tbl-obslog}\firstuse{Tab\ref{tbl-obslog}}
and shown in
Figure \ref{fig-fov}\firstuse{Fig\ref{fig-fov}}.
Two low-gain observations were obtained in each filter.
These observations were secured as
part of the {\sl HST} Cycle 5 program GTO/WFC 6282 (PI: Westphal)
and were placed in the public data archive
at the Space Telescope Science Institute on 1996 July 5.
The datasets were recalibrated at the Canadian Astronomy Data Centre
and retrieved electronically by us using a guest account which was
kindly established for KJM.

These WFPC2 observations contain several types of image defects.
Figure \ref{fig-u2pb0103t}\firstuse{Fig\ref{fig-u2pb0103t}}
shows a negative mosaic image of the
U2PB0103T
dataset.
Besides exhibiting normal cosmic ray damage, this 1100-s F555W exposure
also shows
(1) a satellite trail on the WF4 CCD,
(2) an elevated background near the inner corner of the PC1 CCD,
and
(3) shadows are seen on all four CCDs.
The elevated background near the inner-corner of the PC1 CCD
is probably due to stray light
patterns from a bright star just outside of the PC1 field-of-view
(cf. Fig.\ 7.1.a of Biretta, Ritchie, \& Rudloff \cite{biet1995}).
The shadows seen on all four CCDs are indicative of a serious
problem with these observations because the shadows are
generally seen against an elevated background throughout the entire
WFPC2 field-of-view.
This phenomenon is due to light from the bright sun-lit Earth reflecting
off the optical telescope assembly (OTA) baffles and
the secondary mirror supports (``spider'') and into the WFPC2 instrument.
Elevated backgrounds occur when the angle between the Earth and the OTA axis
is $<$25 degrees
(cf. Fig.\ 11.2.a of Biretta \et \cite{biet1995}).
The background ``sky'' brightened significantly during the course
of these observations
(see Figure
\ref{fig-intensity_histograms}\firstuse{Fig\ref{fig-intensity_histograms}})
indicating that
the {\sl Hubble Space Telescope} experienced earthrise
during these WFPC2 observations of the Ursa Minor dwarf spheroidal galaxy.

The experimental design of these WFPC2 observations was nearly identical
to that of the Carina dwarf spheroidal program GTO/WFC 5637 (PI: Westphal)
which was analyzed by Mighell (\cite{mighell1997}).
We therefore planned to follow Mighell's Carina
photometric reduction procedures in this investigation of the
Ursa Minor dwarf spheroidal.
Unfortunately, the standard cosmic-ray removal procedure failed
spectacularly due to earthrise causing the background sky level to change
rapidly.
We had to improvise more complicated analysis techniques
than ones used by Mighell in his Carina study
in order to obtain stellar photometry of comparable quality.

We found stellar candidates on cosmic-ray cleaned images
which were suitable for the detection of point sources but
unsuitable for further photometric analysis.
The cosmic rays were removed by using the {\sc crrej} task of the
{\sc iraf}
{\sc stsdas.hst\_calib.wfpc} package
with the sky subtraction parameter set to {\sc sky=mode}
instead of the default value of {\sc sky=none} ---
this unusual option was required because
the sky levels of the observations did not scale with exposure time.
We
used {\sc crrej}
to make a clean F555W observation of 2100 s
from the
U2PB0102T
and
U2PB0103T
datasets
and a clean F814W observation of 2300 s
from the
U2PB0105T
and
U2PB0106T
datasets.
Figure\ \ref{fig-crrej}\firstuse{Fig\ref{fig-crrej}}
shows that this procedure repaired most of the cosmic-ray
damage seen in Figure\ \ref{fig-u2pb0103t}.
This procedure is clearly not perfect since
traces of the satellite trial are still visible.
The {\sc sky=mode}
option produces cosmic-ray cleaned images with modal pixel values near zero.
Many background pixels will thus
have negative values which implies
negative background-flux values.
Such physically unrealistic background data values
are quite rightly rejected by many standard CCD stellar photometry
packages.

Unsharp mask images of the clean F555W
and F814W observations were made
using the {\sc{lpd}} (low-pass difference) digital filter
which was designed by Mighell
to optimize the detection of faint stars in {\sl HST} WF/PC and WFPC2
images (Appendix A of Mighell \& Rich \cite{miri1995}, and references therein).
The F555W unsharp mask image
(see Figure\ \ref{fig-crrej_lpd}\firstuse{Fig\ref{fig-crrej_lpd}})
and the F814W unsharp mask image
were then added together to create a
master unsharp mask image of each WF CCD.
A simple peak detector algorithm was then used on the master unsharp
images to create a
list of point source candidates
with coordinates $60 \leq x \leq 790$ and $60 \leq y \leq 790$
on each WF CCD.
This allowed the use of almost the entire field-of-view of each WF camera
while avoiding edge-effects in the outer regions.
We only present the analysis of data obtained from the WF cameras
in this paper.

\newpage
The data were analyzed with
the {\sc ccdcap}\footnote{
IRAF implementations of {\sc ccdcap} are now available over the Wide World Web
at the following site:~
http://www.noao.edu/staff/mighell/ccdcap/
}
digital circular aperture
photometry code developed by Mighell
to analyze
{\sl{HST}} WFPC2
observations
(Mighell \et \cite{mighell1997}, and references therein).
A fixed aperture with a radius of 2.5 pixels was used for all stars
on the WF CCDs.
The local background level was determined from a robust estimate
of the mean intensity value of all pixels between 2.5 and 6.0 pixels
from the center of the circular stellar aperture.
Point source candidates were rejected if either
(1) the measured signal-to-noise ratio of
either instrumental magnitude was SNR$<$10\,;
or (2) the center of the aperture [which was allowed to move in order to
maximize the SNR\,] changed by more than 1.8 pixels
from its detected position on the master unsharp mask.
The Charge Transfer Effect was
removed from the instrumental magnitudes
by using a 4\% uniform wedge
along the Y-axis of each CCD as described in
Holtzman \et (\cite{hoet1995b}).
We used the standard WFPC2 magnitude system
(Holtzman \et \cite{hoet1995b})
which is defined using 1\arcsec\ diameter apertures.
We measured the stars with a smaller aperture (0.5\arcsec\ diameter)
in order to optimize the measured stellar signal-to-noise ratios; usage of
1\arcsec\ diameter apertures resulted in significantly poorer photometry
for the faint stars.
The instrumental magnitudes,
$v_r$ and $i_r$,
were transformed to Johnson $V$ and Cousins $I$ magnitudes
using the following equations
$
V
=
v_r
+ \Delta_r
+ \delta_r
+ [-0.052\!\pm\!0.007](\vmi)
+ [0.027 \!\pm\!0.002](\vmi)^2
+ [21.725\!\pm\!0.005]
$
and
$
I
=
i_r
+ \Delta_r
+ \delta_r
+ [-0.062\!\pm\!0.009](\vmi)
+ [0.025\!\pm\!0.002](\vmi)^2
+ [20.839\!\pm\!0.006]
$
where an instrumental magnitude of zero is defined as one DN s$^{-1}$ at
the high gain state ($\sim$14 e$^-$ DN$^{-1}$\,).
The constants come from
Table 7 of Holtzman \et (\cite{hoet1995b}).
The values for average aperture corrections\footnote{
Observed WFPC2 point spread functions (PSFs)
vary significantly
with wavelength, field position, and time (Holtzman \et \cite{hoet1995a}).
There were not enough bright isolated stars in these WFPC2 observations
to adequately measure the variation of the point spread function
across each WF CCD using the observations themselves.
We measured artificial point spread functions synthesized by the
{\sc{tiny tim }}{\small\sc{version 4.4}}
software package (Krist \cite{kr1993}, Krist \& Hook \cite{krho1997})
to determine the aperture corrections,
$\Delta_r$,
required to convert instrumental magnitudes
measured with an aperture of radius 2.5 pixels
to a standard aperture of radius 5.0 pixels (1\arcsec\ diameter).
A catalog of 289 synthetic point spread functions of a G-type star
was created with
a $17 \times 17$ square grid
for each
filter (F555W and F814W)
and CCD (WF2, WF3, and WF4).
The spatial resolution of one synthetic PSF
every 50 pixels in $x$ and $y$ allowed for the determination
of aperture corrections
for any star in the entire WFPC2 field-of-view to have a
spatial resolution of $\lesssim$35 pixels.
}
, $\langle \Delta_r \rangle$, for each filter/CCD combination
are listed in
Table \ref{tbl-delta}\firstuse{Tab\ref{tbl-delta}}.
The zero-order (``breathing'') aperture corrections\footnote{
Spacecraft jitter during exposures and small focus
changes caused by the {\sl HST} expanding and contracting (``breathing'')
once every orbit are another two important causes of variability in
observed WFPC2 point spread functions.
These temporal variations of WFPC2 PSFs
can cause small, but significant, systematic
offsets in the photometric zeropoints when small apertures are used.
Fortunately,
these systematic offsets can be easily calibrated away by
simply measuring
bright isolated stars on each CCD twice: once with the small aperture
and again with a larger aperture.
The robust mean magnitude difference
between the large and small apertures
is then the
zero-order aperture correction,
$\delta_r$,
for the small aperture which, by definition,
can be positive or negative.
Zero-order aperture corrections
are generally small for long exposures,
however, they
can be quite large for short exposures that were obtained while the WFPC2 was
slightly out of focus (by a few microns)
due to the expansion/contraction of the {\sl HST} during
its normal breathing cycle.
}
for these observations
($\delta_r$ : see
Table \ref{tbl-sdelta}\firstuse{Tab\ref{tbl-sdelta}})
were computed using a large aperture with a radius of 3.5 pixels
and a background annulus of
$3.5\leq r_{\rm sky}\leq7.0$ pixels.

Two ($V,I$) datasets pairs,
(U2PB0102T, U2PB0105T)
and
(U2PB0103T, U2PB0106T),
were reduced independently using
{\sc ccdcap}
and the resulting instrumental magnitudes were transformed to
Johnson $V$ and Cousins $I$ magnitudes.
We determined which objects probably had acceptable photometry
from these independent measurements.
The $V$ measurements of a star,
$V_1$ [$\Leftarrow$ (U2PB0102T, U2PB0105T)\,]
and
$V_2$ [$\Leftarrow$ (U2PB0103T, U2PB0106T)\,]\,,
with photometric errors,
$\sigma_{V_1}$ and $\sigma_{V_2}$,
were determined to be acceptable if the following condition was true:
$
| V_1 - V_2 |
\leq
\max\left(
\left[ 3\sqrt{2}\min(\sigma_{V_1}, \sigma_{V_2}) \right],
0.06
\right)
$\,.
If the condition was satisfied,
we then adopted the quantity,
$\max(V_1,V_2)
- 2.5\log\left[
(1+\left\{10^{0.4}\right\}^{|V_1-V_2|})/2
\right]
$, as the $V$ magnitude of the star and adopted the quantity,
$\sqrt{ ( \sigma^{2}_{V_1}+\sigma^{2}_{V_2} )/2 }$,
as a conservative estimate of its $V$ photometric error, $\sigma_V$.
We assumed that cosmic rays would be the primary cause of
poor photometry and therefore
adopted the photometry of the faintest measurement
of the star whenever the acceptability condition failed.
The adopted $I$ magnitude and $I$ photometric error, $\sigma_I$,
was determined from both $I$ measurements,
$I_1$ [$\Leftarrow$ (U2PB0102T, U2PB0105T)\,]
and
$I_2$ [$\Leftarrow$ (U2PB0103T, U2PB0106T)\,]\,,
in an analogous manner.
Figure \ref{fig-delta_mag}\firstuse{Fig\ref{fig-delta_mag}}
shows the outlier measurements we have identified in this manner.
Figure \ref{fig-cmd_preliminary}\firstuse{Fig\ref{fig-cmd_preliminary}}
gives our preliminary $V$ versus $V\!-I$ color-magnitude diagram CMD of the
observed stellar field in Ursa Minor dwarf spheroidal galaxy.

We present our WFPC2
stellar photometry of 696
stars in the central region of the
Ursa Minor dwarf spheroidal galaxy in
Table\ \ref{tbl-umi_vi_photometry}\firstuse{Tab\ref{tbl-umi_vi_photometry}}.
The first column gives the identification (ID) of the star.
The second and third columns
give the $V$ magnitude and its rms ($1\,\sigma$) photometric error $\sigma_V$.
Likewise, the fourth and fifth columns give the $\vmi$ color and its
rms ($1\,\sigma$) photometric error $\sigma_{(V\!-I)}$.
The sixth column gives the quality flag value of the star.
We only present photometry of stars with signal-to-noise ratios
SNR$\,\geq\,$10 in both the F555W and F814W filters.

\section{DISCUSSION}

\subsection{Color-Magnitude Diagram}

The $V$ versus $\vmi$ color-magnitude diagram of the
observed stellar field in Ursa Minor is shown in
Figure \ref{fig-cmd}\firstuse{Fig\ref{fig-cmd}}.
This CMD
features a sparsely populated blue horizontal branch,
a steep thin red giant branch,
and a narrow subgiant branch.
The main sequence reaches $\sim$2 magnitudes below the
turnoff of the main stellar population of the Ursa Minor galaxy.

Figure \ref{fig-cmd} shows a small amount of foreground
contamination by foreground stars in our Galaxy.
Ratnatunga \& Bahcall (\cite{ratnatunga_bahcall_1985})
used the Bahcall and Soneira Galaxy model
(Bahcall \& Soneira
\cite{bahcall_soneira_1980},
\cite{bahcall_soneira_1984};
Bahcall \et \cite{bahcall_etal_1985})
to predict that 2.3 foreground stars brighter than $V=25$ mag
would be found in one square arcmin in the direction of
Ursa Minor.
Our observation surveys 4.44 arcmin$^2$
of Ursa Minor and we would therefore
expect, from the prediction of Ratnatunga and Bahcall,
to find $\sim$10 foreground stars brighter than $V=25$ mag in
our color-magnitude diagrams.
A direct check with observations is provided by
Figure 2 of Kleyna \et (\cite{kleyna_etal_1998}) which
indicates that while foreground contamination towards
Ursa Minor is small it can not be ignored.
The 4 bright blue stars near  $V\approx20$ mag with colors $(\vmi)<0.3$
mag will be shown below to be probable Ursa Minor horizontal branch stars.
There are a few fainter blue stars seen in Fig.\ \ref{fig-cmd} which
are within $\lea$2 magnitudes of the main-sequence turnoff of the
main Ursa Minor stellar population.
Determining whether these ``blue stragglers'' are actually
members of the Ursa Minor galaxy or are simply Galactic foreground stars
is beyond the scope of this paper.

\subsection{Fiducial Sequence}

The robust median $V-\!I$ color as a function of $V$ magnitude of the
Ursa Minor main sequence, subgiant branch, and base of the red giant branch
($21.5 \leq V \leq 25.0$ mag)
is listed in
Table \ref{tbl-umi_fiducial}\firstuse{Tab\ref{tbl-umi_fiducial}}
and shown in
Figure \ref{fig-cmd_fiducials}\firstuse{Fig\ref{fig-cmd_fiducials}}.
The robust median $V-\!I$ color of a given $\Delta V = 0.2$ mag data subsample
was determined after $\gea$2.4$\sigma$ outliers were iteratively rejected in
5 iterations of a robust fiducial sequence algorithm
(see Appendix A) recently developed by Mighell.
The data in Table \ref{tbl-umi_fiducial} is given in
intervals of $\Delta V = 0.1$ mag.
Since a sampling of $\Delta V=0.2$ was used to determine the robust
median $V-\!I$ colors, we see that there are actually
{\em two} realizations of the
Ursa Minor fiducial sequence given in Table \ref{tbl-umi_fiducial} since
only {\em every other row} in that table represents an
{\em independent measurement} of the true Ursa Minor fiducial sequence.
The first fiducial sequence
is given at $V_{\rm UMi}=21.6, 21.8, \ldots, 24.8$ mag
in Table \ref{tbl-umi_fiducial}
and is shown with {\em open diamonds} in Figure \ref{fig-cmd_fiducials}.
The second fiducial sequence
is given at $V_{\rm UMi}=21.7, 21.9, \ldots, 24.9$ mag
in Table \ref{tbl-umi_fiducial}
and is shown with {\em open squares} in Figure \ref{fig-cmd_fiducials}.

We compare the Ursa Minor fiducial sequences (Table \ref{tbl-umi_fiducial})
with those of the ancient metal-poor Galactic globular cluster M92
(Table A1 in Appendix A) in
Figure \ref{fig-cmd_fiducials}.
We get an excellent fit of the Ursa Minor fiducial sequences to
the M92 fiducial sequences when we
make the M92 fiducial sequence fainter
by $\Delta V = 4.60$ mag and add a small color offset
of $\Delta (V-\!I) = 0.01$ mag.
We show below that the fit is so good
that these fiducials are statistically equivalent
over a 3 magnitude range ($22.0 \leq V_{\rm UMi} < 25.0$ mag)
from the base of the red giant branch of Ursa Minor to $\sim$1.7 magnitudes
below its main-sequence turnoff.
This suggests that the ancient metal-poor Galactic globular cluster M92
is an excellent stellar population analog for the median stellar
population of the Ursa Minor dwarf spheroidal galaxy.
It would not be surprising if the M92 analogy
weakens sometime in the future when
deeper observations with smaller photometric scatter
are analyzed --- especially if these future observations survey a
significantly larger fraction of the Ursa Minor galaxy.

\subsection{$\Delta V_{\rm UMi-M92}$ and $\Delta (\vmi)_{\rm UMi-M92}$}

The $V$ magnitude offset,
$\Delta V_{\rm UMi-M92}$,
and the color offset,
$\Delta (V-\!I)_{\rm UMi-M92}$,
between the Ursa Minor dwarf spheroidal galaxy and the Galactic
globular cluster M92 may be determined by comparing
our fiducial sequences of Ursa Minor (Table 5) and M92 (Table A1).
The parameter space may be investigated through
the application of the following chi-square statistics:
\begin{equation}
\chi_{22.2}^2
\equiv
\sum_{j=1}^{14}
\frac{
\left[
(\vmi)\!\!\raisebox{-1em}{\scriptsize{UMi}}\hspace{-1em}(V_j)
-
(\vmi)^{\prime}\!\!\raisebox{-1em}{\scriptsize{M92}}\hspace{-1em}(V_j-
\Delta V_{\rm UMi-M92})
-
\Delta(\vmi)_{\rm UMi-M92}
\right]^2
}{
\left[
\sigma\!\raisebox{-1em}{\scriptsize{UMi}}\hspace{-1em}(V_j)
\right]^2
+
\left[
\sigma^{\prime}\!\!\raisebox{-1em}{\scriptsize{M92}}\hspace{-1em}(V_j-\Delta
V_{\rm UMi-M92} )
\right]^2
}
\end{equation}
where
$V_j\equiv22.0+0.2j$ mag
and
\begin{equation}
\chi_{22.1}^2
\equiv
\sum_{k=1}^{15}
\frac{
\left[
(\vmi)\!\!\raisebox{-1em}{\scriptsize{UMi}}\hspace{-1em}(V_k)
-
(\vmi)^{\prime}\!\!\raisebox{-1em}{\scriptsize{M92}}\hspace{-1em}(V_k-
\Delta V_{\rm UMi-M92})
-
\Delta(\vmi)_{\rm UMi-M92}
\right]^2
}{
\left[
\sigma\!\raisebox{-1em}{\scriptsize{UMi}}\hspace{-1em}(V_k)
\right]^2
+
\left[
\sigma^{\prime}\!\!\raisebox{-1em}{\scriptsize{M92}}\hspace{-1em}(V_k-\Delta
V_{\rm UMi-M92} )
\right]^2
}
\end{equation}
where
$V_k \equiv 21.9 + 0.2k$ mag.
The color errors are approximated as
\begin{equation}
\sigma
\approx
\frac{1.25\,{\rm adev}}{\sqrt{n}}
\end{equation}
where adev is the average deviation
(column 3 of Tables 5 and A1)
and $n$ is the number of stars in the subsample
(column 6 of Tables 5 and A1).
We use cubic spline interpolations wherever the
M92 fiducial sequence (Table A1) does not have a tabulated value at
$V$ magnitude values of
$V_j-\Delta V_{\rm UMi-M92}$ mag
and
$V_k-\Delta V_{\rm UMi-M92}$ mag.
Usage of cubic spline interpolations is denoted by
prime superscripts over the appropriate terms in the definitions of these
chi-square statistics.

We now use these chi-square statistics to determine the $V$ magnitude
and $\vmi$ color offset between Ursa Minor and M92.
Tables
\ref{tbl-rchisq14}\firstuse{Tab\ref{tbl-rchisq14}}
and
\ref{tbl-rchisq15}\firstuse{Tab\ref{tbl-rchisq15}}
give the reduced chi-square values $\chi_{22.2}^2/14$ and $\chi_{22.1}^2/15$,
respectively,
using $V$ magnitude offsets of
$4.400\leq\Delta V_{\rm UMi-M92}\leq4.800$ mag
and
color offsets of
$\mbox{-0.010}\leq\Delta(\vmi)_{\rm UMi-M92}\leq0.030$ mag.
The residuals of individual fits
(see footnotes a--i in Tables \ref{tbl-rchisq14} and \ref{tbl-rchisq15})
are shown in
Figure \ref{fig-residuals}\firstuse{Fig\ref{fig-residuals}}.

Tables
\ref{tbl-rchisq14}
and
\ref{tbl-rchisq15}
indicate that
a color offset of $\Delta (\vmi)_{\rm UMi-M92}=\mbox{+0.010}$ mag
always produces the lowest reduced chi-square value
--- at any given $V$ magnitude offset.
This is clearly seen in Figure \ref{fig-residuals}.
The residuals systematically become more negative as the
color offset is increased from -0.01 to +0.03 mag;
the residual scatter is minimized (the best fits occur) at +0.01 mag.
We have thus established that the color offset between Ursa Minor and M92
is approximately +0.01 mag.

The top-center panel of
Figure \ref{fig-residuals}
shows that a $V$ magnitude offset of
$\Delta V_{\rm UMi-M92}=4.5$ mag
and a $\vmi$ color offset of
$\Delta (\vmi)_{\rm UMi-M92}=\mbox{+0.010}$ mag
gives systematically large
{\em positive} residuals in the range
$22 \leq V \lea 23$ mag.
This poor fit in the subgiant branch region of Ursa Minor
indicates that the UMi SGB is systematically
{\em fainter } than the shifted M92 SGB.
We have thus established a lower limit of the
$V$ magnitude offset between Ursa Minor and M92:
$\Delta V_{\rm UMi-M92} \gea 4.5$ mag.

The bottom-center panel of
Figure \ref{fig-residuals}
shows that using offsets of
$\Delta V_{\rm UMi-M92}=4.7$ mag
and
$\Delta (\vmi)_{\rm UMi-M92}=\mbox{+0.010}$ mag
gives systematically large
{\em negative} residuals in the range
$22 \leq V \lea 23$ mag.
This poor fit in the subgiant branch region of Ursa Minor
indicates that the UMi SGB is systematically
{\em brighter} than the shifted M92 SGB.
We have thus established an upper limit of the
$V$ magnitude offset between Ursa Minor and M92:
$\Delta V_{\rm UMi-M92} \lea 4.7$ mag.

The 90\%, 95\%, and 99\% confidence limits
($\chi_{22.2}^2/14\,$: 1.50, 1.69, and 2.08)
of the fits given in Table \protect\ref{tbl-rchisq14} are shown in
Figure \ref{fig-contour_rchisq14}\firstuse{Fig\ref{fig-contour_rchisq14}}.
Table \ref{tbl-rchisq14} shows that fits assuming a $V$ magnitude offset
of $\Delta V_{\rm UMi-M92}=\mbox{4.575}$ mag produce the smallest
reduced chi-square value for any given $\vmi$ color offset.
This is clearly seen in Figure \ref{fig-contour_rchisq14}
where the confidence contours are widest at the same $V$ magnitude offset.

The 90\%, 95\%, and 99\% confidence limits
($\chi_{22.1}^2/15\,$: 1.48, 1.66, and 2.04)
of the fits given in Table \protect\ref{tbl-rchisq15} are shown in
Figure \ref{fig-contour_rchisq15}\firstuse{Fig\ref{fig-contour_rchisq15}}.
Table \ref{tbl-rchisq15} shows that fits assuming a $V$ magnitude offset
of $\Delta V_{\rm UMi-M92}=\mbox{4.625}$ mag produce the smallest
reduced chi-square value for any given $\vmi$ color offset.
This is clearly seen in Figure \ref{fig-contour_rchisq15}
where the confidence contours are widest at the same $V$ magnitude offset.

A conservative analysis of Figure \ref{fig-contour_rchisq15}
yields a determination that the
$V$ magnitude offset for the Ursa Minor dSph galaxy from
the Galactic globular cluster M92 is
$\Delta V_{\rm{UMi-M92}} = 4.60\pm0.03$ mag
with 99\% confidence limits of
$4.500\leq\Delta V_{\rm{UMi-M92}}\leq4.700$ mag.
Similarly, a conservative estimate of
the $\vmi$ color offset between Ursa Minor and M92 is
$\Delta (\vmi)_{\rm{UMi-M92}} = 0.010\pm0.005$ mag
with 99\% confidence limits of
${\rm-0.005}\leq\Delta V_{\rm{UMi-M92}}\leq0.020$ mag.

Figure \ref{fig-cmd_with_fiducial}\firstuse{Fig\ref{fig-cmd_with_fiducial}}
shows our Ursa Minor color-magnitude diagram
with the addition of the M92 fiducial sequence of
Johnson \& Bolte (\protect\cite{jobo1998})
which has been plotted with a $V$ magnitude offset of
4.6 mag
and a $V-I$ color offset of 0.01 mag.
We see that the
4 bright blue stars near  $V\approx20$ mag with colors $(\vmi)<0.3$
mag lie underneath the shifted M92 blue horizontal branch;
these stars are probable Ursa Minor horizontal branch stars.
The brighter part of the Ursa Minor red giant branch
($V<22$ mag -- where our fiducials were not compared)
is seen to be slightly redder than the M92 red giant
branch.  This could be evidence that
Ursa Minor is slightly more metal-rich than M92 --- however we
caution the reader not to over-interpret such a small sample
of Ursa Minor red giants.  The current observations can not rule out
that the main stellar population of Ursa Minor has the same metallicity
as M92.

\subsection{Distance, Reddening, and Age of UMi}

With an accurate estimate of $\Delta V_{\rm UMi-M92}$ in hand,
we are now able to determine the apparent $V$ distance modulus of the
Ursa Minor dSph galaxy if we know the apparent $V$ distance modulus
of M92:
$(m-M)_{V}^{\rm UMi}
\equiv
(m-M)_{V}^{\rm M92}
+
\Delta V_{\rm UMi-M92}
$.
The above analysis suggests that the uncertainty in the
the $V$ magnitude offset between Ursa Minor and M92 is small
($\sim$0.03 mag).
This implies that the largest source of
uncertainty in the value of $(m-M)_{V}^{\rm UMi}$ will probably
be the error associated with apparent $V$ distance modulus of M92 itself.
Pont \et (\cite{poet1998}) recently estimated
$(m-M)_{V}^{\rm M92} = 14.67 \pm 0.08$ mag
from their analysis of {\sl Hipparcos} subdwarf parallaxes.
A conservative estimate of the apparent $V$ distance modulus
of the Ursa Minor dSph is then
$(m-M)_{V}^{\rm UMi}
=
(14.67 + 4.60) \pm (0.08 + 0.03)
=
19.27 \pm 0.11
$ mag.

Let us now assume that the $\vmi$ color offset between M92 and Ursa Minor
is completely due to reddening.  The difference in
$\bmv$ reddening between M92 and Ursa Minor would then be
$
\Delta E(\bmv)_{\rm UMi-M92}
=
\Delta (\vmi)_{\rm UMi-M92}/1.3
=
0.008\pm0.004
$ mag
assuming that
$E(\vmi) \approx 1.3\,E(\bmv)$
(Dean, Warren, \& Cousins \cite{dean_warren_cousins_1978}).
Adopting a $\bmv$ reddening for M92,
$E(\bmv)_{\rm M92} = 0.02\pm0.01$ mag
(e.g., Stetson \& Harris \cite{stha1988},
Bolte \& Hogan \cite{bolte_hogan_1995}),
we now determine the $\bmv$ reddening for Ursa Minor to be
$E(\bmv)_{\rm UMi}
=
E(\bmv)_{\rm M92}
+
\Delta E(\bmv)_{\rm UMi-M92}
=
0.03\pm0.01$
mag.
The absorption in V is determined to be
$A_V^{\rm{UMi}} = 0.09\pm0.03$
assuming that
$A_V = 3.1\,E(\bmv)$
(Savage \& Mathis \cite{sama1979}).
Our new $\bmv$ reddening estimate for UMi agrees well with previous estimates
in the literature:
0.03 mag (Zinn \cite{zi1981})
and
$0.02_{-0.02}^{+0.03}$ mag
(Nemec, Wehlau, \& de Oliveira \cite{neet1988}).

Reddening estimates based on
{\sl{COBE/DIRBE}}
and
{\sl{IRAS/ISSA}}
data give $E(\bmv)$ values of
0.023$\pm$0.003 mag and 0.033$\pm$0.004 mag at the respective
positions of the Ursa Minor\footnote{
Estimate derived at the Galactic longitude and latitude
of $({\em{l,b)}}_{\rm{UMi}}=(105\fdg00,44\fdg85)$.
} dwarf spheroidal galaxy and the Galactic globular cluster
M92\footnote{
Estimate derived at the Galactic longitude and latitude
of $({\em{l,b)}}_{\rm{M92}}=(68\fdg34,34\fdg86)$.
}
(Schlegel \et \cite{schlegel_etal_1998}).
The difference between these two values,
$\Delta{\rm{E}}(\bmv)_{\rm{UMi-M92}} = 0.010\pm0.005$ mag,
agrees well with our own estimate of the difference in $\bmv$\ reddening
($0.008\pm0.004$ mag)
which we determined above with a completely different method
(fiducial-sequence fitting).

King \et (\cite{king_etal_1998}) recently suggested that the
$\bmv$\ reddening of M92 may be
$0.04$--$0.05$ mag
greater than canonical values:
${\rm{E}}(\bmv)_{\rm{M92}} = 0.06$--$0.07$ mag.
Reid \& Gizis (\cite{reid_gizis_1998})
observed that the standard $\bmv$\ reddening estimate of M92,
$E(\bmv)_{\rm{M92}} = 0.02$ mag,
is confirmed by Schlegel \et (see above paragraph);
they also note that high reddening estimate of
King \et
is at odds with other studies.
Our determination of the $\bmv$\ reddening difference
between UMi and M92 could be consistent with the high reddening estimate
of King \et
only if the $\bmv$\ reddening of UMi is also $0.04$--$0.05$ greater than
canonical values.  Thus while it is true that reddening is patchy across
the sky, it is rather unlikely that
both M92 and UMi have {\em{exactly the same
amount of extra reddening}} beyond that
predicted from maps of infrared dust emission.
We have thus adopted the traditional $\bmv$\ reddening estimate for
M92 for the sake of consistency with
Schlegel \et (\cite{schlegel_etal_1998})
and older studies of Galactic extinction
(e.g., Burstein \& Heiles \cite{buhe1982}).

We calculate the distance modulus of the Ursa Minor dwarf spheroidal
galaxy to be
$(m-M)_0^{\rm UMi} = 19.18 \pm 0.12$
based on
$(m-M)_{V}^{\rm M92} = 14.67 \pm 0.08$ mag
(Pont \et \cite{poet1998})
which was derived assuming
$\mbox{E}(\bmv)_{\rm M92} = 0.02$ mag
and
$\feh_{\rm M92}=-2.2$ dex
(cf.\ Caretta \& Gratton \cite{cagr1997}, Zinn \& West \cite{ziwe1984}).
Decreasing the adopted $\bmv$ reddening for M92 by 0.01 mag
decreases the distance modulus estimate by 0.02 mag
and increasing the metallicity for M92 by 0.1 dex
increases the distance modulus by 0.03 mag
(Pont \et \cite{poet1998}).

Our new distance estimate for Ursa
Minor is in good agreement with previous determinations based on
early CCD observations in the 1980's once earlier estimates are
placed on the same distance scale.
For example,
Cudworth, Olszewski, \& Schommer (\cite{cuolsc1986})
derived a distance modulus for Ursa Minor,
$(m-M)_0^{\rm{UMi}} = 19.0\pm0.1$ mag,
based on a sliding fit to M92.
They also got the same value from their measurement of the V
magnitude of the horizontal branch at the RR Lyrae gap,
$V_{\rm{RR}}=19.7$ mag, their absorption value,
$A_V^{\rm{UMi}} = 0.1$ mag, and the assumption that the absolute
$V$ magnitude of the RR Lyraes is $M_V^{\rm{RR}}=0.6$ mag.
Harris (\cite{harris_1996}) gives the $V$ magnitude of the
horizontal branch of M92 as $V_{\rm{HB}}^{\rm{M92}} = 15.10$ mag.
With our $V$ magnitude offset value between Ursa Minor and M92,
we expect that the $V$ magnitude of the Ursa Minor horizontal
is $V_{\rm{HB}}^{\rm{UMi}} = 19.70\pm0.03$ mag which exactly agrees
with the measurement of
Cudworth \et (\cite{cuolsc1986}).
Our distance modulus estimate for Ursa Minor
implies that the absolute visual magnitude of the horizontal branch
(at a metallicity of $\feh=-2.2$ dex) is
$M_V^{\rm{HB}}
=
V_{\rm{HB}}^{\rm{UMi}}
-
(m-M)_V^{\rm{UMi}}
=
0.43\pm0.12$ mag
which is consistent with the
Lee, Demarque, \& Zinn (\cite{lee_demarque_zinn_1990}, hereafter LDZ)
distance scale value
$M_{V,{\rm{LDZ}}}^{\rm{RR}}
=
0.17\feh + 0.82
=
0.45$ mag
assuming, of course, that
$M_V^{\rm{HB}}
\approx
M_V^{\rm{RR}}$.
Placing the Ursa Minor distance modulus estimate of
Cudworth \et (\cite{cuolsc1986})
on the LDZ distance scale and assuming our $V$
absorption value,
$A_V^{\rm{UMi}} = 0.09\pm0.03$,
we get a revised estimate of
$(m-M)_0^{\rm{UMi}} = 19.16\pm0.11$ mag
which is just 0.02 mag lower than our own estimate.

How old is the main stellar population of Ursa Minor?
We have shown that the ancient metal-poor Galactic globular cluster M92
is an excellent stellar population analog for the median stellar
population of the Ursa Minor dwarf spheroidal galaxy.
Continuing further with the M92 analogy, we propose that
Ursa Minor and M92 are coeval.
The determination of the
age of the main population of Ursa Minor
reduces then to the problem of determining the age of M92.
The Harris \et (\cite{harris_etal_1997}) analysis of
the Galactic globular clusters NGC 2419 and M92
found that while the full impact of {\sl{Hipparcos}} data
and improving stellar models has yet to be felt,
an age range of 12--15 Gyr for the most metal-poor
Galactic globular clusters is well supported by the
current mix of theory and observations.
Last year, Pont \et (\cite{poet1998}) estimated that M92 is 14 Gyr based on
their analysis of the luminosities of cluster turnoff and subgiant branch
stars.  They noted that their age estimate for M92 should probably be
reduced by $\sim$1 Gyr if diffusion
is important in the cores of globular cluster stars.
Our above analysis used the Pont \et (\cite{poet1998}) distance to M92,
and so we now adopt their age estimate for M92.
Using the M92 analogy one last time, we conclude that
the age of the main stellar
population of the Ursa Minor dwarf spheroidal galaxy
is $\sim$14 Gyr old.

\section{SUMMARY}

The findings of this paper can be summarized as follows:
\begin{itemize}
\item
Our comparison of the fiducial sequence of the Ursa Minor dwarf spheroidal
galaxy with the Galactic globular cluster M92 (NGC 341) indicates that
that the median stellar population of the UMi dSph galaxy
is metal poor
(${\rm[Fe/H]}_{\rm UMi}
\approx
{\rm[Fe/H]}_{\rm M92}
\approx{\rm-2.2}$ dex)
and ancient
($age_{\rm UMi}
\approx
age_{\rm M92}
\approx14$ Gyr).
\item
The $V$ magnitude offset and $\vmi$ color offset between
Ursa Minor and M92 are estimated to be
$\Delta V_{\rm{UMi-M92}} = 4.60\pm0.03$ mag
and $\Delta (\vmi)_{\rm{UMi-M92}} = 0.010\pm0.005$ mag.
\item
The Ursa Minor
$\bmv$ reddening and the absorption in $V$ are estimated to be
E$(\bmv)=0.03\pm0.01$ mag
and
$A_V^{\rm UMi} = 0.09\pm0.03$ mag
assuming that the $\bmv$ reddening for M92 is $0.02\pm0.01$ mag.
\item
We have determined that the distance modulus of the Ursa Minor
dwarf spheroidal galaxy is
$(m-M)_0^{\rm UMi}
=
(m-M)_0^{\rm M92}
+
\Delta V_{\rm{UMi-M92}}
-
A_V^{\rm UMi}
=
19.18 \pm 0.12$ mag
based on the
the adoption of the apparent $V$ distance modulus for M92
of $(m-M)_V^{\rm M92} = 14.67\pm0.08$ mag (Pont \et \cite{poet1998}).
The Ursa Minor dwarf spheroidal galaxy is then at a distance of
$69\pm4$ kpc from the Sun.
\end{itemize}

These {\sl{HST}} observations indicate that Ursa Minor has had a
very simple star formation
history consisting mainly of a single major burst
of star formation about 14 Gyr ago
which probably lasted $\lea$2 Gyr.
While we may have missed minor younger stellar populations
due to the small field-of-view of the WFPC2 instrument,
these observations clearly show that most of the
stars in the central region
Ursa Minor dwarf spheroidal galaxy are ancient.
If the ancient Galactic globular clusters, like M92, formed
concurrently with the early formation of Milky Way galaxy itself,
then the Ursa Minor dwarf spheroidal is probably as old as the Milky Way.

\acknowledgments

We would like to thank Sylvia Baggett for helping us understand
the cause of all the image defects we encountered in these archival images.
We thank the anonymous referee whose comments and suggestions have improved
this article.  We wish to thank Don VandenBerg for bringing to our attention
the article on the distance to NGC 6397 by Reid \& Gizis
which appeared while we were finishing the manuscript.
KJM was supported by a grant from
the National Aeronautics and Space Administration (NASA),
Order No.\ S-67046-F, which was awarded by
the Long-Term Space Astrophysics Program (NRA 95-OSS-16).
Figure \protect\ref{fig-fov}\ was created with an image from the
Digitized Sky Survey\footnote{
Based on photographic data obtained using The UK Schmidt Telescope.
The UK Schmidt Telescope was operated by the Royal Observatory
Edinburgh, with funding from the UK Science and Engineering Research
Council, until 1988 June, and thereafter by the Anglo-Australian
Observatory.  Original plate material is copyright (c) the Royal
Observatory Edinburgh and the Anglo-Australian Observatory.  The
plates were processed into the present compressed digital form with
their permission.  The Digitized Sky Survey was produced at the Space
Telescope Science Institute under US Government grant NAG W-2166.
}.
This research has made use of
NASA's Astrophysics Data System Abstract Service
and the NASA/IPAC Extragalactic Database (NED)
which is operated by the Jet Propulsion Laboratory at
the California Institute of Technology, under
contract with NASA.

\appendix

\section{A ROBUST FIDUCIAL-SEQUENCE ALGORITHM}

Johnson \& Bolte (\cite{jobo1998}, hereafter JB98)
recently published a $V$ versus $V\!-I$ fiducial
sequence for the ancient Galactic globular cluster M92
which is shown in
Figure \ref{fig-A01}\firstuse{Fig\ref{fig-A01}}\
on top of their stellar photometry
which was kindly provided to use by Jennifer Johnson.
JB98 found that mean and mode fitting proved to
be susceptible to outliers due to not having enough stars to
form a strong ridge line in some areas of the color-magnitude
diagram; their M92 fiducial sequence was determined from the best
measured stars and was subsequently drawn by hand and eye.
We now demonstrate that, given enough stars, it is possible to obtain similar
results with a new robust fiducial-sequence algorithm which we present
herein.

The median value of a normal (a.k.a. Gaussian)
distribution is the mean value of
the distribution.
The mean value, ($\bar{x}$),
of a {\em{small}} nearly-normally-distributed sample
is sensitive to the presence of outlier data values;
the median value is less sensitive to outliers and is
therefore considered to be a more robust statistic than the mean.
Likewise, the average deviation (a.k.a.\ mean deviation),
$a \equiv \frac{1}{N} \sum_{i=1}^{N} |x_i-\bar{x}|$,
of a nearly-normally-distributed sample is,
by definition,
less sensitive to outliers than the standard deviation,
$\sigma \equiv [ \frac{1}{N-1} \sum_{i=1}^{N} (x_i-\bar{x})^2 ]^{1/2}$,
of the sample.
The average deviation of a normal distribution is $\sim$0.8 times
the standard deviation of the distribution\footnote{
The average deviation of a normal distribution with a mean of zero and
a standard deviation $\sigma$ is
\begin{displaymath}
a
=
\int_{-\infty}^\infty
|x|
\left[
\frac{1}{\sigma\sqrt{2\pi}}
e^{\displaystyle{-x^2}/{(2\sigma^2)}}
\right]
\,dx
=
\sigma\sqrt{\frac{2}{\pi}}\,.
\end{displaymath}
A normal distribution with a standard deviation of one ($\sigma\!\equiv\!1$),
in the limit of an infinite number of observations,
would thus have an average deviation of $a = \sqrt{2/\pi} \approx 0.7989$\,.
}
\ and approximately 98\% of a normal distribution is
found within 3.0 average deviations of the mean of the distribution\footnote{
\begin{displaymath}
\int_{-3.0a}^{+3.0a}
\left[
\frac{1}{\sigma\sqrt{2\pi}}
e^{\displaystyle{-x^2}/{(2\sigma^2)}}
\right]
\,dx
\approx
\int_{-2.4\sigma}^{+2.4\sigma}
\left[
\frac{1}{\sigma\sqrt{2\pi}}
e^{\displaystyle{-x^2}/{(2\sigma^2)}}
\right]
\,dx
\approx
0.9836\,.
\end{displaymath}
}.

A robust estimate of the mean of a nearly-normally-distributed sample
can be determined by deriving the median of a subsample of the
parent sample that is within 3.0 average deviations of the
median of the parent sample.  This process can, of course,
be repeated until the difference between the parent median and the
subsample median is negligibly small.
Five iterations will generally suffice for the determination of fiducial
sequences from high-quality stellar photometry.

We now apply this algorithm (with 5 iterations)
to the M92 $V-\!I$ color photometry as a function of $V$ magnitude
in order to determine its fiducial sequence:
$[ V_{\rm{M}}, (V-\!I)_{\rm{M}} ]$.
The algorithm results with 0.2-mag slices in $V$ are given in tabular
form in
Table A1\firstuse{TabA1}
and graphically in
Figures
\ref{fig-A02}\firstuse{Fig\ref{fig-A02}},
\ref{fig-A03}\firstuse{Fig\ref{fig-A03}}
and
\ref{fig-A04}\firstuse{Fig\ref{fig-A04}}.

We see that our
M92 fiducial sequence
(Table A1)
matches the fit-by-eye fiducial sequence of JB98 near the
main-sequence turnoff region ($18 \leq V \leq 21$)
to a remarkable degree with a mean
and rms difference of just 0.0004$\pm$0.0047 mag.
The scatter increases slightly for
stars brighter than $V \approx 18$ which is not at all
surprising given the small sample sizes
present on the subgiant branch and red-giant branch of M92
[see column 6 of Table A1].
At the faintest magnitudes ($V>21$ mag) on the main-sequence,
Figures
\ref{fig-A03}
and
\ref{fig-A04}
indicate that our
fiducial sequence M92
is slightly redder than that of JB98.
Noting that the numbers of the stars in the sample
gradually decreases below $V\approx21$
even though the M92 stellar luminosity function
is known to be increasing over this magnitude range
(see, e.g., Stetson \& Harris \cite{stha1988}), we see that
completeness effects become increasingly significant for the JB98 data
below $V\approx21$ mag.  The well-known tendency for faint stars to be
measured too bright explains why the algorithm gave redder $V-I$ colors
than the fit-by-eye values of Johnson \& Bolte who consciously
compensated for this effect
in their determination of the M92 fiducial sequence
(see discussion in \S 3.\ of JB98).

\newpage
\begin{table}
\dummytable\label{tbl-obslog}
\ifundefined{showtables}{
}\else{
  \vspace*{-30mm}
  \hspace*{-20mm}
  \epsfxsize=8.0truein
  \epsfbox{mighell.tab01.eps}
}
\fi
\end{table}

\newpage
\begin{table}
\dummytable\label{tbl-delta}
\ifundefined{showtables}{
}\else{
  \vspace*{-30truemm}
  \hspace*{-20truemm}
  \epsfxsize=8.0truein
  \epsfbox{mighell.tab02.eps}
}
\fi
\end{table}

\newpage
\begin{table}
\dummytable\label{tbl-sdelta}
\ifundefined{showtables}{
}\else{
  \vspace*{-30truemm}
  \hspace*{-20truemm}
  \epsfxsize=8.0truein
  \epsfbox{mighell.tab03.eps}
}
\fi
\end{table}

\newpage
\begin{table}
\dummytable\label{tbl-umi_vi_photometry}
\ifundefined{showtables}{
}\else{
  \vspace*{-30truemm}
  \hspace*{-20truemm}
  \epsfxsize=8.0truein
  \epsfbox{mighell.tab04.eps}
}
\fi
\end{table}

\newpage
\begin{table}
\dummytable\label{tbl-umi_fiducial}
\ifundefined{showtables}{
}\else{
  \vspace*{-30truemm}
  \hspace*{-20truemm}
  \epsfxsize=8.0truein
  \epsfbox{mighell.tab05.eps}
}
\fi
\end{table}

\newpage
\begin{table}
\dummytable\label{tbl-rchisq14}
\ifundefined{showtables}{
}\else{
  \vspace*{-30truemm}
  \hspace*{-20truemm}
  \epsfxsize=8.0truein
  \epsfbox{mighell.tab06.eps}
}
\fi
\end{table}

\newpage
\begin{table}
\dummytable\label{tbl-rchisq15}
\ifundefined{showtables}{
}\else{
  \vspace*{-30truemm}
  \hspace*{-20truemm}
  \epsfxsize=8.0truein
  \epsfbox{mighell.tab07.eps}
}
\fi
\end{table}

\newpage
\begin{table}
\dummytable\label{tbl-A01}
\ifundefined{showtables}{
}\else{
  \vspace*{-30truemm}
  \hspace*{-30truemm}
  \epsfxsize=8.0truein
  \epsfbox{mighell.tabA01.eps}
}
\fi
\end{table}

\begin{table}
\dummytable\label{tbl-x}
\end{table}

\def\fig01cap{
\label{fig-fov}
\noteforeditor{Print this figure ONE (1) COLUMN wide.\newline}
Digitized Sky Survey image of the Ursa Minor dwarf spheroidal galaxy
with the outlines indicating the
{\sl{Hubble Space Telescope}} WFPC2 observation
(see Table \protect\ref{tbl-obslog}).
The entire field shown subtends 10\arcmin\ on a side.
The orientation is North to the top and East to the left.
}
\ifundefined{showfigs}{
  \newpage
  \centerline{{\Large\bf{Figure Captions}}}
  \smallskip
  \figcaption[]{\fig01cap}
}\else{
  \clearpage
  \newpage
  \begin{figure}[p]
    \figurenum{1}
    \vspace*{+0mm}
    \hspace*{+0mm}
    \epsfxsize=6.5truein

    \epsfbox{mighell.fig01.full.eps}
    \vspace*{+0mm}
    \caption[]{\baselineskip 1.15em \fig01cap}
  \end{figure}
}
\fi

\def\fig02cap{
\label{fig-u2pb0103t}
\noteforeditor{Print this figure ONE (1) COLUMN wide.\newline}
A negative mosaic image of the
U2PB0103T
dataset.
}
\ifundefined{showfigs}{
  \figcaption[]{\fig02cap}
}\else{
  \clearpage
  \newpage
  \begin{figure}[p]
    \figurenum{2}
    \vspace*{+0mm}
    \hspace*{+0mm}
    \epsfxsize=6.5truein

    \epsfbox{mighell.fig02.full.eps}
    \vspace*{+0mm}
    \caption[]{\baselineskip 1.15em \fig02cap}
  \end{figure}
}
\fi

\def\fig03cap{
\label{fig-intensity_histograms}
\noteforeditor{Print this figure ONE (1) COLUMN wide.\newline}
The faint portion of the intensity histograms of the
WF3 region $[60:790,60:790]$
of the datasets given in Table \protect\ref{tbl-obslog}.
The background ``sky'' brightened significantly with time.
}
\ifundefined{showfigs}{
  \figcaption[]{\fig03cap}
}\else{
  \clearpage
  \newpage
  \begin{figure}[p]
    \figurenum{3}
    \vspace*{-1.0truein}
    \hspace*{+0mm}

    \epsfxsize=6.5truein
    \epsfbox{mighell.fig03.eps}
    \vspace*{-1.5truein}
    \caption[]{\baselineskip 1.15em \fig03cap}
  \end{figure}
}
\fi

\def\fig04cap{
\label{fig-crrej}
\noteforeditor{Print this figure ONE (1) COLUMN wide.\newline}
The clean combined F555W image
created with the {\sc stsdas crrej} task.
}
\ifundefined{showfigs}{
  \figcaption[]{\fig04cap}
}\else{
  \clearpage
  \newpage
  \begin{figure}[p]
    \figurenum{4}
    \vspace*{+0mm}
    \hspace*{+0mm}
    \epsfxsize=6.5truein

    \epsfbox{mighell.fig04.full.eps}
    \vspace*{+0mm}
    \caption[]{\baselineskip 1.15em \fig04cap}
  \end{figure}
}
\fi

\def\fig05cap{
\label{fig-crrej_lpd}
\noteforeditor{Print this figure ONE (1) COLUMN wide.\newline}
The unsharp mask image of Fig.\ \protect\ref{fig-crrej}.
This image was
created with the {\sc{lpd}} (low-pass difference) digital filter
(Mighell \& Rich \protect\cite{miri1995}).
}
\ifundefined{showfigs}{
  \figcaption[]{\fig05cap}
}\else{
  \clearpage
  \newpage
  \begin{figure}[p]
    \figurenum{5}
    \vspace*{+0mm}
    \hspace*{+0mm}
    \epsfxsize=6.5truein

    \epsfbox{mighell.fig05.full.eps}
    \vspace*{+0mm}
    \caption[]{\baselineskip 1.15em \fig05cap}
  \end{figure}
}
\fi

\def\fig06cap{
\label{fig-delta_mag}
\noteforeditor{Print this figure ONE (1) COLUMN wide.\newline}
Comparison of the $VI$ photometry based on the independent magnitude
measurements $(V_1,\-I_1)$ and $(V_2,\-I_2)$ which were derived from the
respective observations (U2PB0102T,\-U2PB0105T) and (U2PB0103T,\-U2PB0106T).
The {\em gray}\,({\em{green}}) {\em{squares}} indicate stars that have
good photometry in both observations.  The {\em black}\,({\em{blue}})
{\em{circles}} show probable outlier values which have poor photometry
in at least one of the observations.
}
\ifundefined{showfigs}{
  \figcaption[]{\fig06cap}
}\else{
  \clearpage
  \newpage
  \begin{figure}[p]
    \figurenum{6}
    \vspace*{+0mm}
    \hspace*{+35mm}
    \epsfxsize=3.25truein
    \epsfbox{mighell.fig06.eps}
    \vspace*{-0.0truein}
    \caption[]{\baselineskip 1.15em \fig06cap}
  \end{figure}
}
\fi

\def\fig07cap{
\label{fig-cmd_preliminary}
\noteforeditor{Print this figure ONE (1) COLUMN wide.\newline}
The preliminary $V$ versus $V\!-I$ color-magnitude diagram
of the observed stellar field in Ursa Minor dwarf spheroidal galaxy.
The 503
{\em black}\,({\em{blue}})\,{\em{squares}}
are objects with good photometry in all four observations.
The 219
{\em gray}\,({\em{green}})\,{\em{diamonds}}
are objects where at least one of the observations was
flagged as a probable outlier in Figure
\protect\ref{fig-delta_mag}.
The 17 objects overlayed with \S\ symbols are probable galaxies;
the 9 $+$\ symbols indicate hot pixels.
}
\ifundefined{showfigs}{
  \figcaption[]{\fig07cap}
}\else{
  \clearpage
  \newpage
  \begin{figure}[p]
    \figurenum{7}
    \vspace*{-1.0truein}
    \hspace*{+0mm}
    \epsfxsize=6.5truein
    \epsfbox{mighell.fig07.eps}
    \vspace*{-1.5truein}
    \caption[]{\baselineskip 1.15em \fig07cap}
  \end{figure}
}
\fi

\def\fig08cap{
\label{fig-cmd}
\noteforeditor{Print this figure ONE (1) COLUMN wide.\newline}
The $V$ versus $V\!-I$ color-magnitude diagram
of the Ursa Minor dwarf spheroidal galaxy.
The error bars indicate\ rms ($1\,\sigma$) uncertainties
for a single star at the corresponding magnitude.
}
\ifundefined{showfigs}{
  \figcaption[]{\fig08cap}
}\else{
  \clearpage
  \newpage
  \begin{figure}[p]
    \figurenum{8}
    \vspace*{-1.0truein}
    \hspace*{+0mm}
    \epsfxsize=6.5truein
    \epsfbox{mighell.fig08.eps}
    \vspace*{-1.5truein}
    \caption[]{\baselineskip 1.15em \fig08cap}
  \end{figure}
}
\fi

\def\fig09cap{
\label{fig-cmd_fiducials}
\noteforeditor{Print this figure ONE (1) COLUMN wide.\newline}
An expanded version of Figure \protect\ref{fig-cmd} with
the two Ursa Minor fiducial sequences
(Table \protect\ref{tbl-umi_fiducial})
shown with {\em open diamonds}
($V_{\rm UMi}=21.6,21.8,\ldots,24.8$ mag)
and {\em open squares}
($V_{\rm UMi}=21.7,21.9,\ldots,24.9$ mag).
We also show, for the sake of comparison, the
M92 fiducial sequence (Table A1 of Appendix A)
which has been plotted ({\em curves from left to right}) assuming
a shift in $V-\!I$ color
(from M92 to UMi) of 0.005, 0.010, and 0.015 mag and a shift in $V$ magnitude
of 4.625, 4.600, 4.575 mag, respectively.
}
\ifundefined{showfigs}{
  \figcaption[]{\fig09cap}
}\else{
  \clearpage
  \newpage
  \begin{figure}[p]
    \figurenum{9}
    \vspace*{-1.0truein}
    \hspace*{+0mm}
    \epsfxsize=6.5truein
    \epsfbox{mighell.fig09.eps}
    \vspace*{-1.5truein}
    \caption[]{\baselineskip 1.15em \fig09cap}
  \end{figure}
}
\fi

\def\fig10cap{
\label{fig-residuals}
\noteforeditor{Print this figure TWO (2) COLUMNS wide.\newline}
The residuals of individual fits from
Tables \protect\ref{tbl-rchisq14} and \protect\ref{tbl-rchisq15}.
The residuals of fits marked with footnotes a--i in those tables
are shown here in panels (a)--(i).
The
{\em black} ({\em{blue}})
residuals are from
Tables \protect\ref{tbl-rchisq14}
and their reduced chi-square values are shown in the bottom-left corner
of each panel.
The
{\em gray} ({\em{cyan}})
residuals are from
Tables \protect\ref{tbl-rchisq15}
and their reduced chi-square values are shown in the bottom-right corner
of each panel.
The assumed $V$ magnitude offset and
$\vmi$ color offset between Ursa Minor and M92
is displayed in the top-right corner of each panel.
}
\ifundefined{showfigs}{
  \figcaption[]{\fig10cap}
}\else{
  \clearpage
  \newpage
  \begin{figure}[p]
    \figurenum{10}
    \vspace*{-1.0truein}
    \hspace*{+0mm}
    \epsfxsize=6.5truein
    \epsfbox{mighell.fig10.eps}
    \vspace*{-1.5truein}
    \caption[]{\baselineskip 1.15em \fig10cap}
  \end{figure}
}
\fi

\def\fig11cap{
\label{fig-contour_rchisq14}
\noteforeditor{Print this figure ONE (1) COLUMN wide.\newline}
The 90\%, 95\%, and 99\% confidence limits of
the fits given in Table \protect\ref{tbl-rchisq14}
are shown, respectively, with {\em dashed, solid, dotted curves}.
The fits associated with the three shifted M92 fiducials shown in
Figure \protect\ref{fig-cmd_fiducials}
are displayed here with
$\ast$ symbols.
Note that all three fits are found within the 90\% confidence limit.
}
\ifundefined{showfigs}{
  \figcaption[]{\fig11cap}
}\else{
  \clearpage
  \newpage
  \begin{figure}[p]
    \figurenum{11}
    \vspace*{-1.0truein}
    \hspace*{+0mm}
    \epsfxsize=6.5truein
    \epsfbox{mighell.fig11.eps}
    \vspace*{-1.5truein}
    \caption[]{\baselineskip 1.15em \fig11cap}
  \end{figure}
}
\fi

\def\fig12cap{
\label{fig-contour_rchisq15}
\noteforeditor{Print this figure ONE (1) COLUMN wide.\newline}
Same as Figure \protect\ref{fig-contour_rchisq14} with the addition of the
90\%, 95\%, and 99\% confidence limits of
the fits given in Table \protect\ref{tbl-rchisq15}
being shown, respectively, with {\bf thick} {\em dashed, solid, dotted curves}.
}
\ifundefined{showfigs}{
  \figcaption[]{\fig12cap}
}\else{
  \clearpage
  \newpage
  \begin{figure}[p]
    \figurenum{12}
    \vspace*{-1.0truein}
    \hspace*{+0mm}
    \epsfxsize=6.5truein
    \epsfbox{mighell.fig12.eps}
    \vspace*{-1.5truein}
    \caption[]{\baselineskip 1.15em \fig12cap}
  \end{figure}
}
\fi

\def\fig13cap{
\label{fig-cmd_with_fiducial}
\noteforeditor{Print this figure ONE (1) COLUMN wide.\newline}
Same as Fig.\ \protect\ref{fig-cmd}
 with the addition of the M92 fiducial sequence of
Johnson \& Bolte (\protect\cite{jobo1998})
which has been plotted with a $V$ magnitude offset of
4.6 mag
and a $V-I$ color offset of 0.01 mag.
Only the Ursa Minor stars in the range
$22\leq V \leq 25$ mag
and M92 stars fainter than $V=17.6$ mag
(see Fig.\ \protect\ref{fig-A01})
were used for the comparison of the fiducial sequences.
}
\ifundefined{showfigs}{
  \figcaption[]{\fig13cap}
}\else{
  \clearpage
  \newpage
  \begin{figure}[p]
    \figurenum{13}
    \vspace*{-1.0truein}
    \hspace*{+0mm}
    \epsfxsize=6.5truein
    \epsfbox{mighell.fig13.eps}
    \vspace*{-1.5truein}
    \caption[]{\baselineskip 1.15em \fig13cap}
  \end{figure}
}
\fi

\def\figA01cap{
\label{fig-A01}
\noteforeditor{Print this figure ONE (1) COLUMN wide.\newline}
The M92 fiducial sequence of
Johnson \& Bolte (\protect\cite{jobo1998}; hereafter JB98)
is shown as the {\em{gray}} ({\em{red}}) {\em{curve}}
on top of their M92 stellar photometry.
The {\em{dark gray}} ({\em{blue}}) {\em{squares}} are the
stars with the best photometry: 2008 out of 3581 stars
had $\geq$14 observations {\em{and}}
{\sc{daophot}} parameter {\sc{chi}} values $\leq$1.3\,.
The {\em{gray}} ({\em{turquoise}}) {\em{squares}} are the
remaining stars with lower quality photometry.
}
\ifundefined{showfigs}{
  \figurenum{A1}
  \figcaption[]{\figA01cap}
}\else{
  \clearpage
  \newpage
  \begin{figure}[p]
    \figurenum{A1}
    \vspace*{-20mm}
    \hspace*{+0mm}
    \epsfxsize=6.5truein
    \epsfbox{mighell.figA01.eps}
    \vspace*{-40mm}
    \caption[]{\baselineskip 1.15em \figA01cap}
  \end{figure}
}
\fi

\def\figA02cap{
\label{fig-A02}
\noteforeditor{Print this figure TWO (2) COLUMNS wide.\newline}
Each panel shows a
$\Delta V = 0.2$ mag wide
subsample of the M92 data of JB98
for $18.1 < V \leq 22.1$ mag in steps of 0.2 mag.
The {\em{dark gray}} ({\em{blue}}) {\em{histogram}} in each panel
shows the $V-I$ color distribution (in steps of 0.005 mag)
of most of the stars within the $V$ magnitude range
shown in the upper-right corner of the panel.
The total number of stars in each panel is
shown in the upper-right corner of the panel
below the $V$ magnitude range.
The {\em{dashed line}} and {\em{4-digit number}}
in each panel shows the median $V-I$ color,
$(V-I)_{\rm{M}}$ [column 2 of Table A1],
of the subsample of stars enclosed within the
{\em{dotted lines}}.
Stars within the dotted lines of each panel all have $V-I$ colors
within 3 adev [column 3 of Table A1] of
$(V-I)_{\rm{M}}$.
A consistency check is provided in the form of the
{\em{gray}} ({\em{green}}) {\em{cumulative fraction distribution}}
of the $V-I$ color distribution for all the stars in each
panel.
}
\ifundefined{showfigs}{
  \figurenum{A2}
  \figcaption[]{\figA02cap}
}\else{
  \clearpage
  \newpage
  \begin{figure}[p]
    \figurenum{A2}

    \vspace*{-40mm}
    \hspace*{-2mm}
    \epsfxsize=3.5truein
    \epsfbox{mighell.figA02.part1of2.eps}
    \hspace*{-13mm}
    \epsfxsize=3.5truein
    \epsfbox{mighell.figA02.part2of2.eps}
    \vspace*{-20mm}
    \caption[]{\baselineskip 1.15em \figA02cap}
  \end{figure}
}
\fi

\def\figA03cap{
\label{fig-A03}
\noteforeditor{Print this figure ONE (1) COLUMN wide.\newline}
The alternating {\em{open squares}} and the {\em{open diamonds}}
plot our M92 fiducial sequence
(Table A1). The M92 fiducial sequence of JB98
is shown as the {\em{gray}} ({\em{magenta}}) {\em{curve}}.
The best M92 stellar photometry of JB98 (defined in Fig.\ A1)
is shown with {\em{dark gray}} ({\em{blue}}) {\em{squares}}
with the remainder shown with
{\em{gray}} ({\em{turquoise}}) {\em{squares}}.
}
\ifundefined{showfigs}{
  \figurenum{A3}
  \figcaption[]{\figA03cap}
}\else{
  \clearpage
  \newpage
  \begin{figure}[p]
    \figurenum{A3}
    \vspace*{-20mm}
    \hspace*{+0mm}
    \epsfxsize=6.5truein
    \epsfbox{mighell.figA03.eps}
    \vspace*{-40mm}
    \caption[]{\baselineskip 1.15em \figA03cap}
  \end{figure}
}
\fi

\def\figA04cap{
\label{fig-A04}
\noteforeditor{Print this figure ONE (1) COLUMN wide.\newline}
The difference between our M92 fiducial sequence
Table A1) and a spline fit of the M92 fiducial sequence of JB98.
The mean and rms difference over the entire $V$ magnitude range is
$0.007\!\pm\!0.014$ mag; near the main-sequence turnoff
($18.0\!\leq\!V\!\leq\!21.0$) the mean and rms difference is significantly
better: $0.0004\!\pm\!0.0047$ mag. The {\em{dark gray}} ({\em{blue}})
{\em{errorbars}}\ show the lower limit estimate of the $1$$\sigma$ errors
[$\sigma\approx(\,1.25\,{\rm{adev}}/\protect\sqrt{n})$
where adev is the average deviation and $n$ is the number stars in the
given subsample (see Table A1).  The {\em{light gray}} ({\em{cyan}})
{\em{errorbars}}\ show the conservative upper limit estimate of the
$1$$\sigma$ errors ($\sigma \approx $1.25\,adev).
}

\ifundefined{showfigs}{
  \figurenum{A4}
  \figcaption[]{\figA04cap}
}\else{
  \clearpage
  \newpage
  \begin{figure}[p]
    \figurenum{A4}
    \vspace*{-20mm}
    \hspace*{+0mm}
    \epsfxsize=6.5truein
    \epsfbox{mighell.figA04.eps}
    \vspace*{-40mm}
    \caption[]{\baselineskip 1.15em \figA04cap}
  \end{figure}
}
\fi

\end{document}